\documentclass[10pt]{article}
\usepackage{fullpage}
\usepackage{amsmath}
\usepackage{amssymb}
\usepackage[dvips]{epsfig}
\usepackage{color}

\def\##1{\underline{#1}}
\def\=#1{\underline{\underline{#1}}}

\def\+#1{\underline{\bf #1}}
\def\*#1{\underline{\underline{\bf #1}}}

\def\r#1{(\ref{#1})}
\def\l#1{\label{#1}}
\def\c#1{\cite{#1}}

\def\le{\left(}
\def\ri{\right)}
\def\les{\left[}
\def\ris{\right]}
\def\lec{\left\{}
\def\ric{\right\}}

\def\.{\mbox{ \tiny{$^\bullet$} }}

\def\eps{\epsilon}

\def\epso{\epsilon_{\scriptscriptstyle 0}}

\def\muo{\mu_{\scriptscriptstyle 0}}

\def\co{c_{\scriptscriptstyle 0}}

\begin{document}

\begin{center}

{\bf {\LARGE Towards a  realization of \vspace{2pt}
\\Schwarzschild-(anti-)de Sitter spacetime \vspace{2pt}
\\ as a particulate metamaterial}}

\vspace{10mm} \large

 Tom G. Mackay\footnote{E--mail: T.Mackay@ed.ac.uk}\\
{\em School of Mathematics and
   Maxwell Institute for Mathematical Sciences\\
University of Edinburgh, Edinburgh EH9 3JZ, UK}\\
and\\
 {\em NanoMM~---~Nanoengineered Metamaterials Group\\ Department of Engineering Science and Mechanics\\
Pennsylvania State University, University Park, PA 16802--6812,
USA}\\
 \vspace{3mm}
 Akhlesh  Lakhtakia\footnote{E--mail: akhlesh@psu.edu}\\
 {\em NanoMM~---~Nanoengineered Metamaterials Group\\ Department of Engineering Science and Mechanics\\
Pennsylvania State University, University Park, PA 16802, USA}\\
and\\
 {\em Materials Research Institute\\
Pennsylvania State University, University Park, PA 16802, USA}

\end{center}

\vspace{4mm}

\normalsize

\begin{abstract}

While metamaterials offer the potential to realize Tamm mediums
which represent vacuous spacetime subjected to gravitational fields,
practical formulations for suitable metamaterials have not hitherto
been developed. This matter is addressed by establishing a
metamaterial formulation for the Tamm medium representing
Schwarzschild-(anti-)de Sitter spacetime. Our formulation is
remarkably simple and does not involve a
 complex nanostructure of the type that is often associated with metamaterials.
Instead it  is based on
 the
homogenization of only isotropic dielectric and isotropic magnetic
component mediums, which are distributed randomly as oriented
spheroidal particles.
Using the inverse Bruggeman homogenization formalism,
 we demonstrated that a  wide range of constitutive
parameter values for the Tamm medium may be accessed through varying
the particle shape, volume  fraction or relative permittivity and
relative permeability of the component mediums. The presented
formulation is appropriate for the regions of spacetime which lie
outside the event horizon for Schwarzschild  spacetime and inside
the event horizon for de Sitter spacetime; there are no such
restrictions for anti-de Sitter spacetime.

\end{abstract}

\vspace{5mm} \noindent  {\bf PACS numbers:}  04.40.Nr, 98.80.-k,
41.20.Jb

\vspace{5mm} \noindent  {\bf Keywords:} metamaterials, Bruggeman
homogenization formalism, Tamm medium, Schwarzschild-(anti-)de
Sitter spacetime

\section{Introduction}

Metamaterials constitute a notable class of engineered materials
which offer opportunities for realizing such exotic phenomenons as
negative refraction and cloaking \c{Metamaterials}. Furthermore,
metamaterials offer unique opportunities to investigate general
relativistic scenarios \c{Smolyaninov_JOP}. This arises from the
formal analogy that exists between light propagation in vacuous
curved spacetime and propagation in a certain nonhomogeneous
anisotropic or bianisotropic medium, known as a Tamm medium
\c{Skrotskii,Plebanski,SS}.
 Lately, theoretical metamaterial-based
analogs of  black holes \c{Smolyaninov_NJP}, de Sitter spacetime
\c{Li_1,Li_2},  strings \c{Li_3} including
 cosmic
strings \c{Spinning_string},  and wormholes \c{Greenleaf},  for
examples,  have been proposed. Crucially, these spacetimes are
amenable to representation by metamaterials because their metrics
are time-independent.

While metamaterials may in principle be exploited to construct
analogs of curved spacetime, concrete details of how this can be
achieved in practice are conspicuously absent from the literature. A
noteworthy exception is a recent description of a metamaterial
representation of an artificial electromagnetic black hole
\c{Lu_JAP}, based upon the homogenization of simple components.
However, the two-dimensional black hole  considered is not
astrophysical. In the following sections we develop a metamaterial
formulation for a rather more complex curved spacetime scenario,
which includes the Schwarzschild black hole and (anti-)de Sitter
spacetime as specializations. Our formulation~---~which is
remarkable for its simplicity~---~relies on
 the
homogenization of isotropic dielectric and isotropic magnetic
component mediums, which are distributed randomly as oriented
spheroidal particles.

As regards notation, 3-vectors are underlined with the $\hat{}$
symbol denoting a unit vector, whereas 3$\times$3 dyadics are double
underlined with $\=I$ being the identity.
 The speed of light in vacuum in the absence of a
gravitational field is $\co = 1/\sqrt{\epso \muo}$, where $\epso =
8.854\times 10^{-12}$~F~m~$^{-1}$ and $\muo = 4\pi\times
10^{-12}$~H~m$^{-1}$.

\section{Tamm medium for
Schwarzschild-(anti-)de Sitter spacetime}

Static Schwarzschild-(anti-)de Sitter spacetime is conventionally
represented by the line element
\c{gh1977,pod1999,car2001}
\begin{equation}
d\tilde{s}^2 = \le 1- \tilde{F} \ri d \tilde{t}^2 - \frac{1}{1-F}\,d
\tilde{r}^2 - \tilde{r}^2 (d\tilde{\theta}^2+\sin^2\tilde{\theta} \,
d\tilde{\phi}^2) , \l{spher}
\end{equation}
expressed in spherical coordinates with
  the adopted signature
$(+,-,-,-)$. Herein the function
\begin{equation}
 \tilde{ F} = \frac{2GM}{\co^2 \tilde{r}} +  \frac{\Lambda \tilde{r}^2}{3 \co^2},
\end{equation}
   with
 $M \ge 0$ being the mass of the corresponding Schwarzschild black hole,  $G$  the
gravitational constant, and $\Lambda$  the cosmological constant.
The spacetime is called Schwarzschild--de Sitter spacetime if the
cosmological constant $\Lambda$ is positive; on the other hand,
 Schwarzschild--anti--de Sitter
spacetime  is characterized by  $\Lambda < 0$.
  \c{exact}. There are two noteworthy specializations, namely:
\begin{itemize}
\item[(a)]  the $\Lambda = 0$ regime
 corresponds to Schwarzschild spacetime, and
\item[(b)] the $M= 0$ regime corresponds to
de Sitter or anti-de Sitter spacetime, according to whether
$\Lambda$ is positive or negative, respectively.
\end{itemize}
In terms of Cartesian coordinates $\tilde{x} = \tilde{r} \sin
\tilde{\theta} \cos \tilde{\phi}$, $\tilde{y} =  \tilde{r} \sin
\tilde{\theta} \sin \tilde{\phi}$, and $\tilde{z} = \tilde{r} \cos
\tilde{\theta} $, the line element \r{spher} is represented by the
metric $\tilde{g}_{\alpha \beta}$ as\footnote{Roman indexes take the
values 1, 2 and 3; while Greek indexes take the values 0, 1, 2, and
3.}
\begin{eqnarray}
\les \, \tilde{g}_{\alpha \beta} \, \ris & = &  \left(
\begin{array}{cccc}
1-\tilde{F} & 0 & 0 & 0\\
\\
0 & -  1 - \frac{\displaystyle  \tilde{F} \tilde{x}^2}{\displaystyle
\tilde{r}^2 \le 1-\tilde{F} \ri }
 & -\frac{\displaystyle  \tilde{F} \tilde{x} \tilde{y}}{\displaystyle \tilde{r}^2 \le 1-\tilde{F} \ri } & -\frac{\displaystyle
   \tilde{F} \tilde{x} \tilde{z}}{\displaystyle \tilde{r}^2 \le 1-\tilde{F} \ri } \\
\\
0 & -\frac{\displaystyle  \tilde{F}\tilde{x}\tilde{y}}{\displaystyle
\tilde{r}^2 \le 1-\tilde{F} \ri }
& -  1 - \frac{\displaystyle  \tilde{F}\tilde{y}^2}{\displaystyle \tilde{r}^2 \le 1-\tilde{F} \ri }
& -\frac{\displaystyle  \tilde{F}\tilde{y}\tilde{z}}{\displaystyle \tilde{r}^2 \le 1-\tilde{F} \ri } \\
\\
0 & -\frac{\displaystyle  \tilde{F}\tilde{x}\tilde{z}}{\displaystyle
\tilde{r}^2 \le 1-\tilde{F} \ri } & -\frac{\displaystyle
\tilde{F}\tilde{y}\tilde{z}}{\displaystyle \tilde{r}^2 \le
1-\tilde{F} \ri } & -
 1
- \frac{\displaystyle  \tilde{F}\tilde{z}^2}{\displaystyle
\tilde{r}^2 \le 1-\tilde{F} \ri }
\end{array} \right). \l{gab}
\end{eqnarray}
 For our purposes here, it is more convenient to
work with a diagonal metric. Therefore, we implement the (spatial)
coordinate transformation represented by
\begin{equation}
\le \begin{array}{cccc} t & x & y & z \end{array} \ri^T = \mathsf{M}
\cdot \le \begin{array}{cccc} \tilde{t} & \tilde{x} & \tilde{y} &
\tilde{z}
\end{array} \ri^T,
\end{equation}
with the change of basis matrix
\begin{equation}
\mathsf{M} = \le
\begin{array}{cccc} 1 &0 &0 &0 \vspace{6pt} \\0& \displaystyle{\frac{\tilde{x}}{\tilde{r}}}
 & -\displaystyle{ \frac{\tilde{z}}{\sqrt{\tilde{x}^2 + \tilde{z}^2}}}
& - \displaystyle{\frac{\tilde{x}\tilde{y}}{\tilde{r}
\sqrt{\tilde{x}^2 + \tilde{z}^2}}} \vspace{6pt}
 \\ 0& \displaystyle{\frac{\tilde{y}}{\tilde{r}}} &0 & \displaystyle{\frac{\sqrt{\tilde{x}^2 + \tilde{z}^2}}{\tilde{r}}}
  \vspace{6pt} \\ 0& \displaystyle{\frac{\tilde{z}}{\tilde{r}}}
  & \displaystyle{\frac{\tilde{x}}{\sqrt{\tilde{x}^2 + \tilde{z}^2}} }& -\displaystyle{ \frac{\tilde{y}\tilde{z}}{\tilde{r}
   \sqrt{\tilde{x}^2 + \tilde{z}^2 }}} \end{array} \ri.
\end{equation}
Since $\mathsf{M}$ is an orthogonal matrix, we have $\tilde{r} \equiv r  = \sqrt{x^2 + y^2 + z^2}$.
With respect to this new coordinate system, the metric has the
diagonal form
\begin{equation}
\les \, g_{\alpha \beta} \, \ris \equiv \mathsf{M}^T \cdot \les \,
\tilde{g}_{\alpha \beta} \, \ris \cdot \mathsf{M} =
\le
\begin{array}{cccc} 1- \tilde{F} & 0 &0 &0 \\
0 & \displaystyle{\frac{1}{\tilde{F} - 1}} & 0 & 0 \\
0 &0 & -1 &0\\
0 &0 &0 & -1
\end{array}
\ri. \l{newg}
\end{equation}

Following the noncovariant approach pioneered by Tamm
\c{Skrotskii,Plebanski,SS}, the electromagnetic response of vacuum
in curved spacetime represented by the metric \r{newg} may be
described by the constitutive relations of an equivalent,
instantaneously responding, medium  (known as a Tamm medium) per
\c{MLS_Schwarz}
\begin{equation}
\left.
\begin{array}{l}
\#D = \epso \=\gamma \. \#E
\\
\#B= \muo \=\gamma \. \#H
\end{array}
\right\}, \l{CR}
\end{equation}
wherein SI units are implemented. Here, $\=\gamma$ is  the
3$\times$3 dyadic equivalent of the metric $\les \, \gamma_{ab} \,
\ris $ with components
\begin{equation}
\gamma_{ab}  = -  \frac{g^{ab}}{g_{00}} .
\end{equation}
That is, the dyadic $\=\gamma$ has the uniaxial form
\begin{equation}
\=\gamma = \mbox{diag} \le 1, \gamma, \gamma \ri,
\end{equation}
where the scalar $\gamma = 1/\le 1 - \tilde{F} \ri$.

In view of our quest to construct a Tamm medium with relative
permittivity dyadic $\=\gamma$ and relative permeability dyadic
$\=\gamma$, it is of interest to consider the spatial dependency of
the parameter $\gamma$. In the case of Schwarzschild spacetime,
$\gamma$ with respect to $r$ exhibits a singularity at $r= 2 G M /
\co^2$, which corresponds to an event horizon. In the case of
(anti-)de Sitter spacetime, $\gamma$ with respect to $r$ exhibits
singularities at $r= \pm \co \sqrt{3 /\Lambda}$. Since $r \ge 0$,
this corresponds to a single event horizon for de Sitter spacetime
at $r=  \co \sqrt{3 /\Lambda}$ and no event horizon for anti-de
Sitter spacetime. Furthermore, $\gamma > 0$ outside the event
horizon for Schwarzschild spacetime but inside the event horizon for
de Sitter spacetime; and $\gamma > 0$ everywhere for anti-de Sitter
spacetime. These features are illustrated in
 Fig.~\ref{F_gamma1}, wherein $\gamma$ is plotted versus $r$ for
Schwarzschild and (anti-)de Sitter spacetimes for $ M = | \Lambda |
= 1.3$ using the normalizations $\co = G =1$.

The situation for Schwarzschild-(anti-)de Sitter spacetime is rather
more complicated, as is shown  in
 Fig.~\ref{F_gamma2} where $\gamma$ is plotted versus $r$ for
Schwarzschild-(anti-)de Sitter spacetimes for $ M = | \Lambda | =
0.3$ using the normalizations $\co = G =1$. Now, for $r \ge 0$,
$\gamma$ with respect to $r$ exhibits two singularities for
Schwarzschild-de Sitter spacetime (and a third one for $r<0$), while
there is also one singularity for Schwarzschild-anti-de Sitter
spacetime. Also, $\gamma > 0$ between the two singularities for
Schwarzschild-de Sitter spacetime and outside the singularity for
Schwarzschild-anti-de Sitter spacetime.

The distinction between the $\gamma > 0$ and $\gamma < 0$ regimes is
of relevance when one considers the phenomenon of electromagnetic
planewave propagation with negative phase velocity, which is
pertinent to metamaterials which support negative refraction
\c{ML_PRB}. In fact, negative phase velocity can arise in the
$\gamma < 0$ regime but not in the $\gamma > 0$ regime
\c{MSL_Cosmo,MLS_Schwarz}. The homogenization procedure presented in
Sec.~\ref{iBr} is suitable for the $\gamma > 0$ regime only.

Finally in this section, we comment on a practical aspect of
realizing the Tamm medium.
 In practice, the nonhomogeneous nature of
the Tamm medium could be catered for by subdividing the space of
interest into local neighbourhoods which are sufficiently small to
be considered approximately homogeneous. The
 inverse homogenization procedures
developed  in Sec.~\ref{iBr} would then be applied locally. This
piecewise  homogeneous approach for the Tamm medium is documented in
detail elsewhere \c{MLS_NJP}.

\section{Inverse Bruggeman formalism} \l{iBr}

The Tamm medium described by the constitutive relations \r{CR} is a
uniaxial dielectric-magnetic medium with identical relative
permittivity and relative permeability dyadics. In order to
construct such a medium, we turn to homogenization using as our
basis the well-established Bruggeman formalism \c{WLM_MOTL,EAB}.

Let us consider the homogenization of four component mediums,
labelled $a$, $b$, $c$ and $d$. Components $a$ and $b$ are isotropic
dielectric mediums with relative permittivities $\eps_a$ and
$\eps_b$, and relative permeabilities $\mu_a = \mu_b =1$. Components
$c$ and $d$ are  isotropic magnetic mediums with relative
permeabilities $\mu_c$ and $\mu_d$, and relative permittivities
$\eps_c = \eps_d =1$. The four component mediums are assumed to be
randomly distributed, with respective volume fractions $f_a$, $f_b$,
$f_c$, $f_d \in \le 0,1 \ri$ with $f_d = 1 - f_a - f_b - f_c$. Each component medium is
composed of spheroidal particles which are small compared to the
electromagnetic wavelengths under consideration. The axis of these
spheroids for all four component mediums is taken to be aligned with
the symmetry axis of $\=\gamma$, namely the $\hat{\#x}$ axis. Thus,
the surface of each spheroid relative to its centre is prescribed by
the vector
\begin{equation}
\#r_{\,s} = \rho_\ell \, \=U_{\, \ell} \cdot \hat{\#r},
\end{equation}
where the positive-definite shape dyadic
\begin{equation}
\=U_{\, \ell} = \mbox{diag} \le 1, U_\ell, U_\ell \ri, \qquad \qquad
\le \ell = a, b, c, d \ri,
\end{equation}
the radial unit vector is $\hat{\#r}$, and $\rho_\ell$ is a linear
measure of size. Spheroids characterized by $U_\ell > 1$ are prolate
whereas  oblate spheroids are characterized by $U_\ell < 1$, and the
 degenerate spherical case corresponds to $U_\ell = 1$.

The Bruggeman formalism provides estimates of the relative
permittivity dyadic $\=\eps_{\,Br} = \mbox{diag} \le \eps^x_{Br},
\eps_{Br}, \eps_{Br} \ri$ and the relative permeability dyadic
$\=\mu_{\,Br}= \mbox{diag} \le \mu^x_{Br}, \mu_{Br}, \mu_{Br} \ri$
of the homogenized composite medium (HCM), as follows. Let us
introduce the polarizability density dyadics
\begin{equation}
\left.
\begin{array}{l}
 \=a^\eps_{\,\ell} = \le \eps_\ell \=I - \=\eps_{\,Br} \ri
\cdot \les \, \=I + \=D^\eps_{\,\ell} \cdot \le \eps_\ell \=I -
\=\eps_{\,Br} \ri \ris^{-1} \vspace{6pt}\\
 \=a^\mu_{\,\ell} = \le \mu_\ell \=I - \=\eps_{\,Br} \ri
\cdot \les \, \=I + \=D^\mu_{\,\ell} \cdot \le \mu_\ell \=I -
\=\mu_{\,Br} \ri \ris^{-1}
\end{array}
\right\} ,\qquad \qquad \le \ell = a, b, c, d \ri.
\end{equation}
  The depolarization dyadics $\=D^{\eps,\mu}_{\, \ell}$ herein are
given by \c{Michel,MW97}
\begin{equation}
\=D^{m}_{\, \ell} = \mbox{diag} \le D^{mx}_{\ell}, D^{m}_{\ell},
D^{m}_{\ell} \ri, \qquad \qquad \le \ell = a, b, c, d; \;\; m =
\eps, \mu \ri,
\end{equation}
where the components
\begin{eqnarray}
D^{mx}_{\ell} &=& \frac{1 - g(\sigma_\ell)}{m_{Br} \le \sigma_\ell - 1 \ri},\\
D^m_\ell &=& \frac{1}{2 m_{Br} U^2_\ell} \le g(\sigma_\ell) - \frac{1 -
g(\sigma_\ell)}{ \sigma_\ell - 1 } \ri,
\end{eqnarray}
with the function
\begin{equation}
g(\sigma_\ell) = \left\{ \begin{array}{lcr}
\displaystyle{\frac{1}{\sqrt{1 - \sigma_\ell}}
\tanh^{-1} \le \sqrt{1 - \sigma_\ell} \ri}, && 0 < \sigma_\ell < 1\\
\displaystyle{\frac{1}{\sqrt{\sigma_\ell - 1}} \tan^{-1} \le \sqrt{
\sigma_\ell -1} \ri}, &&
 \sigma_\ell > 1 \end{array}
\right. ,
\end{equation}
and the dimensionless parameter $\sigma_\ell = U^2_\ell m^x_{Br} /
m_{Br}$. Parenthetically, these depolarization dyadics are only
defined for the $\sigma_\ell
> 0$ regime  \c{Michel}~---~which corresponds the $\gamma > 0$ regime for the Tamm medium.
According to the Bruggeman formalism, the constitutive parameters of
the HCM are related to those of the component mediums by the dyadic
equations \c{WLM_MOTL,EAB}
\begin{equation}
\left.
\begin{array}{ll}
\=A^\eps = \=0 \vspace{4pt}\\
\=A^\mu = \=0
\end{array}
\right\}, \l{A_eq}
\end{equation}
where
\begin{equation}
\=A^m = f_a \,\=a^m_{\,a} + f_b\, \=a^m_{\,b} +  f_c\, \=a^m_{\,c} +
f_d \,\=a^m_{\,d}, \qquad (m = \eps, \mu).
\end{equation}
In fact, as the diagonal dyadics $\=A^m$ have the form $\mbox{diag}
\le A^{mx}, A^m, A^m \ri$, ($m = \eps, \mu$), the dyadic equations
\r{A_eq} contain only four independent  scalar equations, which are
coupled via the constitutive parameters for the HCM.

Conventionally, homogenization formalisms are applied in the forward
sense, wherein the constitutive parameters of the HCM are estimated
from a knowledge of the constitutive parameters of the component
mediums. However, since
 our aim here is to find values of $\eps_{a,b}$, $\mu_{c,d}$,
$f_{a,b,c}$ and $U_{a,b,c,d}$ such that the corresponding HCM
coincides with the Tamm medium specified by the constitutive
relations \r{CR}, we  apply the Bruggeman formalism in its inverse
sense. While formal expressions of the inverse Bruggeman  formalism
have been developed \c{WSW_MOTL}, these formal expressions can be
ill-defined \c{Cherkaev}.\footnote{We note that certain constitutive
parameter regimes
 have been found to be problematic for the inverse Bruggeman homogenization formalism \c{SSJ_TGM}, but these  regimes do not overlap with
 the regimes
 considered here. }  In practice, it is more convenient to
exploit direct numerical methods in order to implement the inverse
formalism \c{ML_JNP}.

 The following three distinct applications of the inverse
Bruggeman formalism are considered, and illustrated using numerical
examples in the next section. In each application, there are four
scalar parameters to be determined.
\begin{itemize}
\item[(i)] Assuming that the relative permittivities $\eps_{a,b}$ and relative permeabilities
 $\mu_{c,d}$ are known, and that $U_a = U_b = U_c = U_d = U$,
 we determine the common shape parameter $U$ and the volume
 fractions $f_a$, $f_b$ and $f_c$.
 \item[(ii)] Assuming that the relative permittivities $\eps_{a,b}$ and relative permeabilities
 $\mu_{c,d}$ are known, and that the volume fractions $f_{a,b,c}$ are fixed,
 we determine the  shape parameters  $U_a$, $U_b$, $U_c$ and $U_d$.
\item[(iii)] Assuming that the shape parameters $U_{a,b,c,d}$ and the volume fractions $f_{a,b,c}$ are fixed,
 we determine  the relative permittivities $\eps_{a,b}$ and relative permeabilities
 $\mu_{c,d}$.
\end{itemize}

As a representative example, let us concentrate on the numerical
implementation of application (i)~---~the numerical  implementations
for applications (ii) and (iii) are analogous. A modified
 Newton--Raphson technique \c{Stark,Kampia}
may be implemented to  extract the  volume fractions $f_{a,b,c}$ and
common shape factor $U$ from  eqs.~\r{A_eq}.  Using this technique,
the solutions at step $k+1$, namely
 $\lec U^{(k+1)}, f_a^{(k+1)},\,f_b^{(k+1)},\,f_c^{(k+1)} \ric$, are derived
 from those at step $k$, namely $\lec U^{(k)}, f_a^{(k)},\,f_b^{(k)},\,f_c^{(k)} \ric$, via
the recursive scheme
\begin{equation}
\left.
\begin{array}{l}
\displaystyle{U^{(k+1)} = U^{(k)} - \frac{A^{\eps x} (  U^{(k)},
f_a^{(k)},\,f_b^{(k)},\,f_c^{(k)} ) }{\frac{\partial}{\partial U}
A^{\eps x}  (  U^{(k)},
f_a^{(k)},\,f_b^{(k)},\,f_c^{(k)} )}} \vspace{6pt} \\
\displaystyle{f_a^{(k+1)} = f_a^{(k)} - \frac{A^\eps  (  U^{(k+1)},
f_a^{(k)},\,f_b^{(k)},\,f_c^{(k)} ) }{\frac{\partial}{\partial f_a}
A^\eps  (  U^{(k+1)},
f_a^{(k)},\,f_b^{(k)},\,f_c^{(k)} )}} \vspace{6pt} \\
\displaystyle{f_b^{(k+1)} = f_b^{(k)} - \frac{A^{\mu x}  (
U^{(k+1)}, f_a^{(k+1)},\,f_b^{(k)},\,f_c^{(k)} )
}{\frac{\partial}{\partial f_b} A^{\mu x}  (  U^{(k+1)},
f_a^{(k+1)},\,f_b^{(k)},\,f_c^{(k)} )}}
\vspace{6pt} \\
\displaystyle{f_c^{(k+1)} = f_c^{(k)} - \frac{A^{\mu }  ( U^{(k+1)},
f_a^{(k+1)},\,f_b^{(k+1)},\,f_c^{(k)} ) }{\frac{\partial}{\partial
f_c} A^{\mu }  (  U^{(k+1)}, f_a^{(k+1)},\,f_b^{(k+1)},\,f_c^{(k)}
)}}
\end{array} \l{NR1}
\right\}.
\end{equation}
wherein the components of the dyadics $\=A^{\eps, \mu}$ are
expressed as functions of the unknown parameters $\lec U,
f_a,\,f_b,\,f_c \ric$.

For convergence of the  scheme \r{NR1}, it is vital that the initial
estimates $\lec U^{(0)}, f_a^{(0)},\,f_b^{(0)},\,f_c^{(0)} \ric$ are
chosen to be sufficiently
 close to the true solution. The forward Bruggeman formalism
 can be utilized in order to generate
 suitable initial estimates, as we now outline.
Let  $\lec \breve{\eps}^x_{Br}, \breve{\eps}_{Br},
\breve{\mu}^x_{Br}, \breve{\mu}_{Br} \ric $ denote the forward
Bruggeman estimates of the HCM's relative permittivity and relative
permeability parameters, computed  for physically reasonable ranges
of the parameters $U$ and $f_{a,b,c}$,
 namely $ U \in
\le U^+, U^- \ri$ and  $f_{a,b,c} \in \le f^+_{a,b,c}, f^-_{a,b,c}
\ri$. Then:
\begin{itemize}
\item[(1)] Fix $ f_a = \le f^-_a + f^+_a \ri /2$, $ f_b = \le f^-_b + f^+_b \ri /2$,
and $ f_c = \le f^-_c + f^+_c \ri /2$. For all values of $U \in \le
U^-, U^+ \ri$, find the value $U^\dagger$  for which the quantity
 \begin{equation} \Delta = \sqrt{\le
\breve{\eps}^x_{Br}  - 1 \ri^2 + \le \frac{\breve{\eps}_{Br} -
\gamma}{\gamma} \ri^2 + \le \breve{\mu}^x_{Br} - 1 \ri^2 + \le
\frac{\breve{\mu}_{Br} - \gamma}{\gamma} \ri^2}
\end{equation} is minimized.
\item[(2)] Fix $U = U^\dagger$,
 $ f_b = \le f^-_b + f^+_b \ri /2$,
and $ f_c = \le f^-_c + f^+_c \ri /2$.
 For all values of $f_a \in \le f_a^-, f_a^+ \ri$, find the
value $f_a^\dagger$  for which $\Delta$ is minimized.
\item[(3)] Fix $U = U^\dagger$, $ f_a = f_a^\dagger$, and $ f_c = \le f^-_c + f^+_c \ri
/2$. For all values of $f_b \in \le f_b^-, f_b^+ \ri$, find the
value $f_b^\dagger$ for which  $\Delta$ is minimized.
\item[(4)] Fix $U = U^\dagger$, $ f_a = f_a^\dagger$, and $ f_b = f_b^\dagger$.
 For all values of $f_c \in \le f_c^-, f_c^+ \ri$, find the value
$f_c^\dagger$ for which  $\Delta$ is minimized.
\end{itemize}
The steps (1)--(4) are repeated, using $f_a^\dagger$, $f_b^\dagger$,
and $f_c^\dagger$ as the fixed values of $f_{a,b,c}$
 in step (i),   $f_b^\dagger$ and  $f_c^\dagger$ as the fixed values of
$f_{b,c}$ in step (ii), and  $f_c^\dagger$  as the fixed value of
$f_{c}$ in step (iii),
 until $\Delta$ becomes sufficiently
small.

In our numerical studies, we found that when $\Delta < 0.01$, the
values of $U^\dagger$, $f_a^\dagger$, $f_b^\dagger$ and
$f_c^\dagger$ provide suitable initial estimates for the modified
Newton-Raphson scheme \r{NR1}.  In fact, we found that this
technique of iteratively scanning the space of possible solutions
could itself be
 used to find the inverse Bruggeman solutions, in some cases with a  faster rate  of
convergence than the Newton-Raphson method and in a manner that is
less sensitive to the initial estimates.

\section{Numerical illustrations}

We present numerical illustrations of the inverse homogenization
applications (i)--(iii) described in Sec.~\ref{iBr}. For each
illustration, the parameter estimates for the component mediums are
calculated as functions of $\gamma$. The ranges $\gamma > 1$ and $0<
\gamma < 1$ are considered~---~the  $\gamma > 1$ range corresponds
to Schwarzschild-de Sitter spacetime whereas the $0 < \gamma < 1$
corresponds to both Schwarzschild-de Sitter  and
Schwarzschild-anti-de Sitter spacetime, depending upon the magnitude
of $\Lambda$.
 For all the  numerical results presented in
 Figs.~\ref{fig3}--\ref{fig5}, the degree of
 convergence of the numerical schemes
 was $< 1 \%$, and for most of the plotted points this value was $< 0.1 \%$.

Let us begin with application (i). The common shape parameter $U$
and the volume fractions $f_{a,b,c}$ are plotted versus $\gamma$ in
Fig.~\ref{fig3}. For $0.8 < \gamma < 0.9$ we chose $\eps_a = 2$,
$\eps_b = 0.1$, $\mu_c = 1.8$ and $\mu_d =0.2$, while for $1 <
\gamma < 2.9$ we chose $\eps_a = 8$, $\eps_b = 0.3$, $\mu_c = 7.4$
and $\mu_d =0.4$. The common shape parameter increases sharply as
$\gamma$ increases, for both $0.8 < \gamma < 0.9$ and $1 < \gamma <
2.9$ ranges, while no particularly noteworthy trend is obvious from
the  volume fraction plots.

In Fig.~\ref{fig4}, the shape parameters $U_{a,b,c,d}$ are presented
as functions of $\gamma$ for the application (ii). For $0.795 <
\gamma < 0.805$ we chose $\eps_a = 2$, $\eps_b = 0.1$, $\mu_c = 1.8$
and $\mu_d =0.2$ (as we chose for Fig.~\ref{fig3} for $0.8 < \gamma
< 0.9$) and $f_{a,b} = 0.21$, $f_c =0.3$ , while for $1.9 < \gamma <
2.2$ we chose $\eps_a = 8$, $\eps_b = 0.3$, $\mu_c = 7.4$ and $\mu_d
=0.4$ (as we chose for Fig.~\ref{fig3} for $1 < \gamma < 2.9$) and
$f_{a,b,c} = 0.25$. We see that $U_{a,c}$ decrease sharply as
$\gamma$ increases for $0.795 < \gamma < 0.805$, but increase
sharply as $\gamma$ increases for $1.9 < \gamma < 2.2$; the opposite
trend is exhibited by $U_{b,d}$.

Finally, for application (iii), plots of the relative permittivities
$\eps_{a,b}$ and  relative permeabilities $\mu_{c,d}$ against
$\gamma$ are provided in Fig.~\ref{fig5}.
 For $0.7 < \gamma < 1$ we chose the common shape parameter $U \equiv U_{a,b,c,d} = 0.01$ and
  the common volume fraction $f_{a,b,c} = 0.25$, while for $1.1 < \gamma < 2.9$
we chose the common shape parameter $U \equiv U_{a,b,c,d} = 5$ and
  the common volume fraction $f_{a,b,c} = 0.25$. For both $0.7 < \gamma < 1$ and $1.1 < \gamma < 2.9$
we found that the numerical schemes provide values of $\eps_a$ and
$\mu_c$ which are almost the same, and values of $\eps_b$ and
$\mu_d$ which are almost the same.

We close this section by remarking that the relative permittivities
and relative permeabilities featured in Figs.~\ref{fig3}--\ref{fig5}
are neither unreasonably high nor unreasonably low from a physical
perspective. Indeed, as regards the smallest values of
$\eps_{a,b,c,d}$ and $\mu_{a,b,c,d}$ featured in
Figs.~\ref{fig3}--\ref{fig5}, we note that materials with relative
permittivities and relative permeabilities close to zero are
currently subjects of intense investigation
 \c{Alu,Lovat,Cia_PRB}.

\section{Closing remarks}

By the homogenization of simple arrangements of four simple
component mediums, the Tamm medium for Schwarzschild-(anti-)de
Sitter spacetime may be constructed. The simplicity of this
 construction is especially noteworthy in view of the complex
 nanostructures which characterize many metamaterials designed for cloaking
  or negative refraction applications \c{Metamaterials}, for example.
 A wide range of constitutive
parameter values for the Tamm medium can be accessed through varying
the particle shape, volume  fraction or relative permittivity and
relative permeability of the component mediums. Thus, the inverse
homogenization formulation delivers a practical strategy for
designing an experimental analog for Schwarzschild-(anti-)de Sitter
spacetime.

While the inverse homogenization scenario chosen for presentation in
Sec.~\ref{iBr} was based on four component mediums, this formulation
is not unique. Indeed, fewer components mediums may be used.  For
example, the Tamm medium for Schwarzschild-(anti-)de Sitter
spacetime  could similarly be conceptualized as arising from the
homogenization of two component mediums, both of which being
isotropic dielectric--magnetic mediums distributed as oriented
spheroids. Alternatively, two component mediums could used which
were both uniaxial dielectric--magnetic mediums (with parallel
symmetry axes) distributed as spherical particles. However, the
four--component formulation presented herein incorporates very
simple components, and accordingly offers a very large degree of
freedom in choosing constitutive parameters.

\begin{figure}[!ht]
\centering
\includegraphics[width=4.9in]{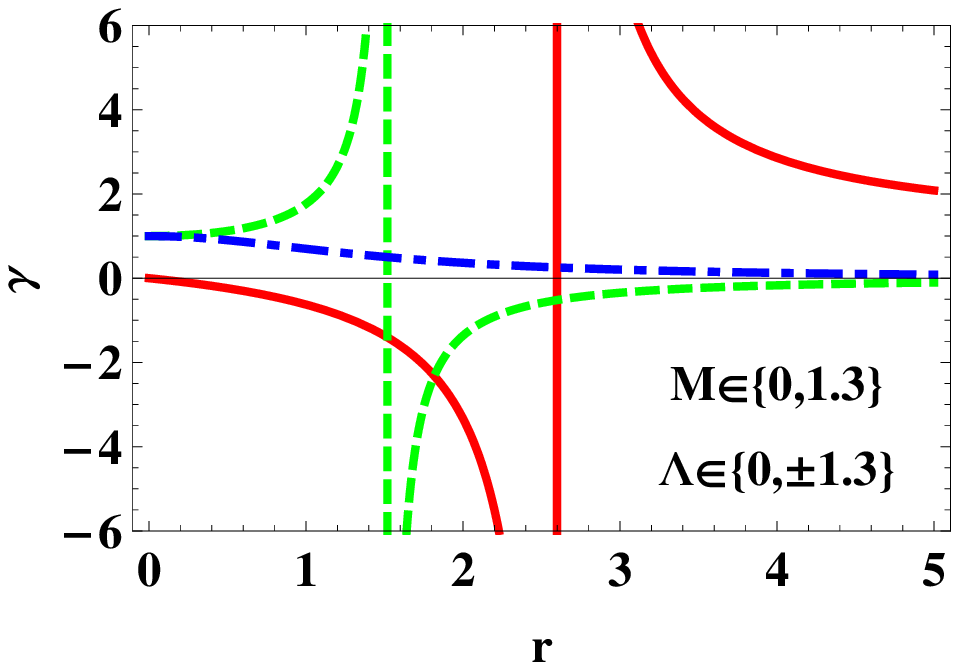}
 \caption{\label{F_gamma1}
 The  parameter $\gamma$ plotted versus $r$ for
Schwarzschild  (solid, red), de Sitter (dashed, green) and anti-de
Sitter (broken dashed, blue) spacetimes. Here $M \in \lec 0, 1.3
\ric$, $\Lambda \in \lec 0, \pm 1.3 \ric$ with the normalization $G
= \co = 1$.
 }
\end{figure}

\begin{figure}[!ht]
\centering
\includegraphics[width=4.9in]{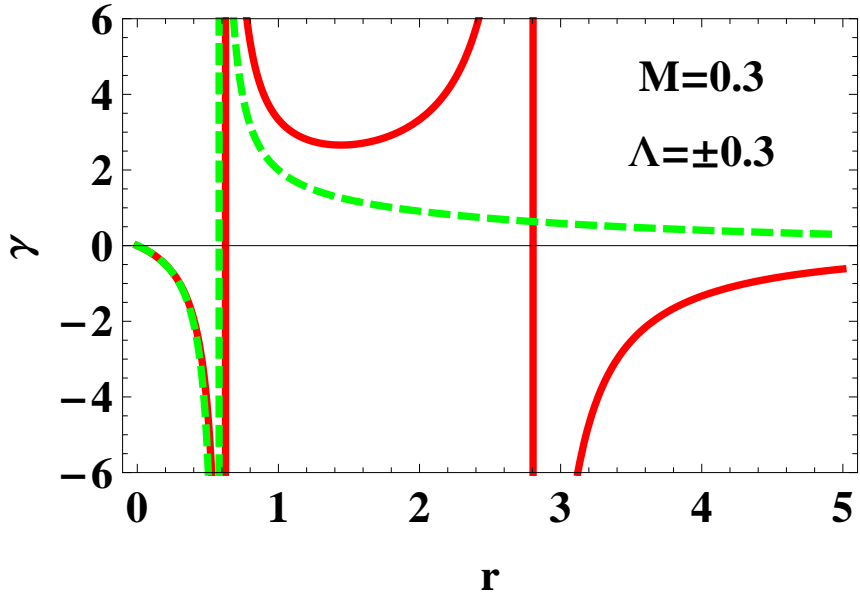}
 \caption{\label{F_gamma2}
 The  parameter $\gamma$ plotted versus $r$ for
Schwarzschild-de Sitter  (solid, red) and Schwarzschild-anti-de
Sitter (dashed, green)  spacetimes. Here $M = 0.3 $, $\Lambda  = \pm
0.3 $ with the normalization $G = \co = 1$.
 }
\end{figure}

\begin{figure}[!ht]
\centering
\includegraphics[width=4.9in]{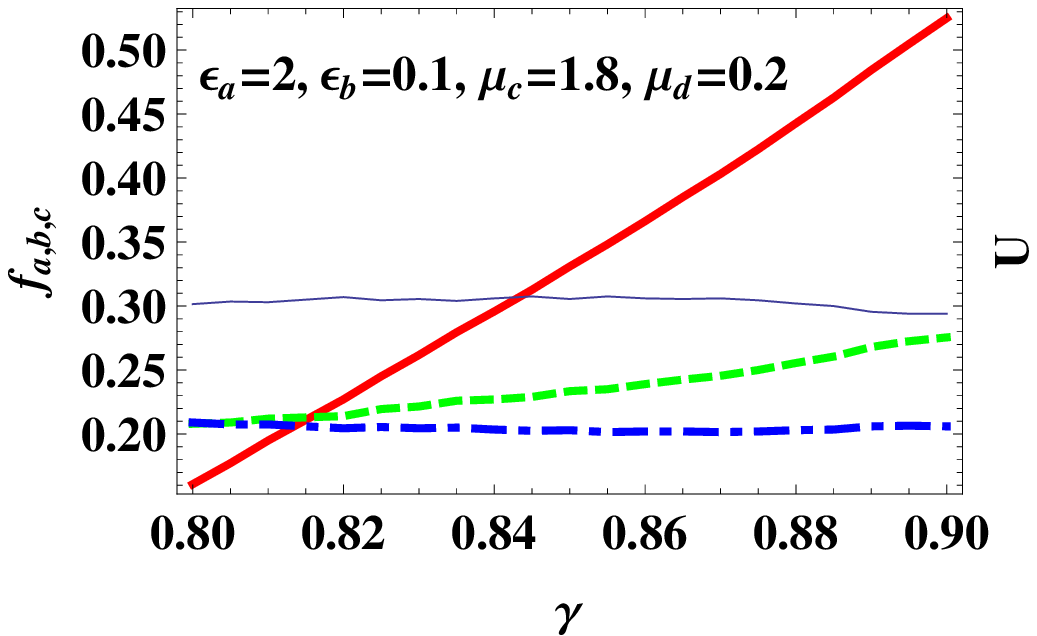}\\
\includegraphics[width=4.9in]{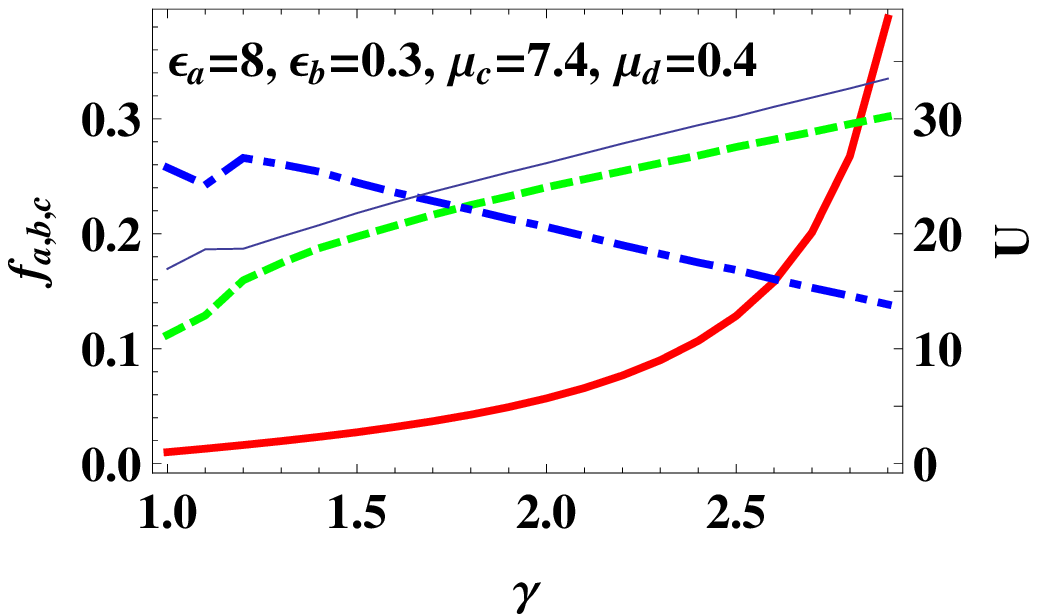}
 \caption{\label{fig3}
 The common shape parameter $U$ (thick solid,  red)  and volume fractions $f_{a}$
  (dashed, green), $f_b$ (broken dashed, blue) and $f_c$ (thin solid, blue)
  plotted versus $\gamma$.  The relative permittivities $\epsilon_a = 2$,
 $\epsilon_b = 0.1$  and relative
 permeabilities $\mu_c = 1.8$, $\mu_d = 0.2$ for $0.8 < \gamma < 0.9$
 (upper);
  and
 $\epsilon_a = 8$, $\epsilon_b = 0.3$, $\mu_c = 7.4$, $\mu_d = 0.4$   for $1 < \gamma < 2.9$ (lower).  }
\end{figure}

\begin{figure}[!ht]
\centering
\includegraphics[width=4.9in]{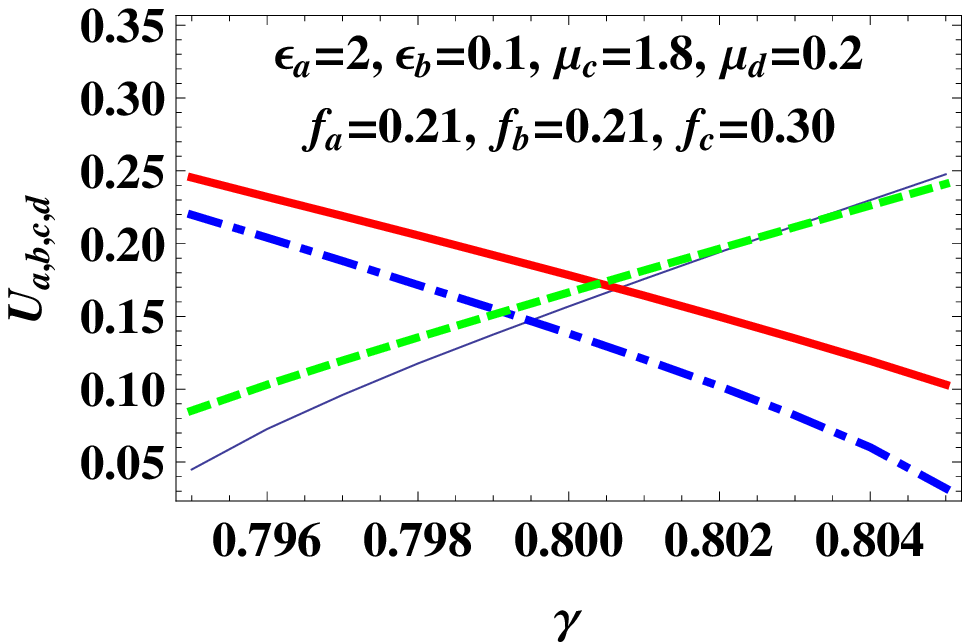}\\
\includegraphics[width=4.9in]{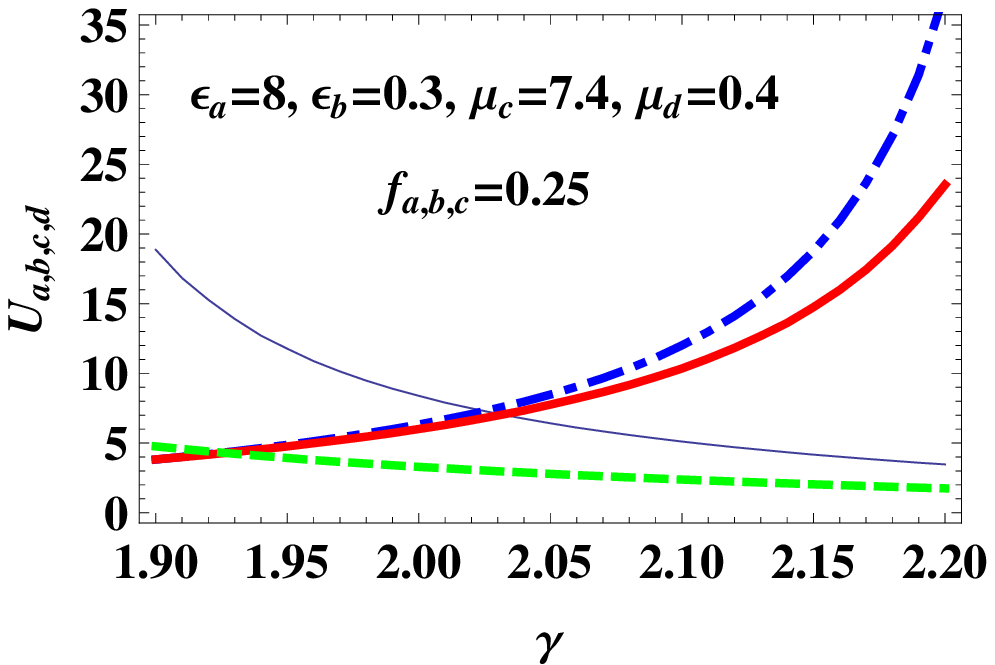}
 \caption{\label{fig4}
  The shape parameters $U_{a}$ (thick solid,  red), $U_{b}$ (dashed, green), $U_{c}$ (broken dashed, blue) and $U_{d}$  (thin solid, blue)
    plotted versus $\gamma$. The relative permittivities $\epsilon_a = 2$, $\epsilon_b = 0.1$ and relative
  permeabilities $\mu_c = 1.8$, $\mu_d = 0.2$
and volume
 fractions $f_a = 0.21$, $f_b = 0.21$ and $f_c = 0.3$
 for $0.795 < \gamma < 0.805$ (upper);
  and
 $\epsilon_a = 8$, $\epsilon_b = 0.3$, $\mu_c = 7.4$, $\mu_d = 0.4$ and $f_{a,b,c,d} = 0.25$  for $1.9 < \gamma < 2.2$ (lower).  }
\end{figure}

\begin{figure}[!ht]
\centering
\includegraphics[width=4.9in]{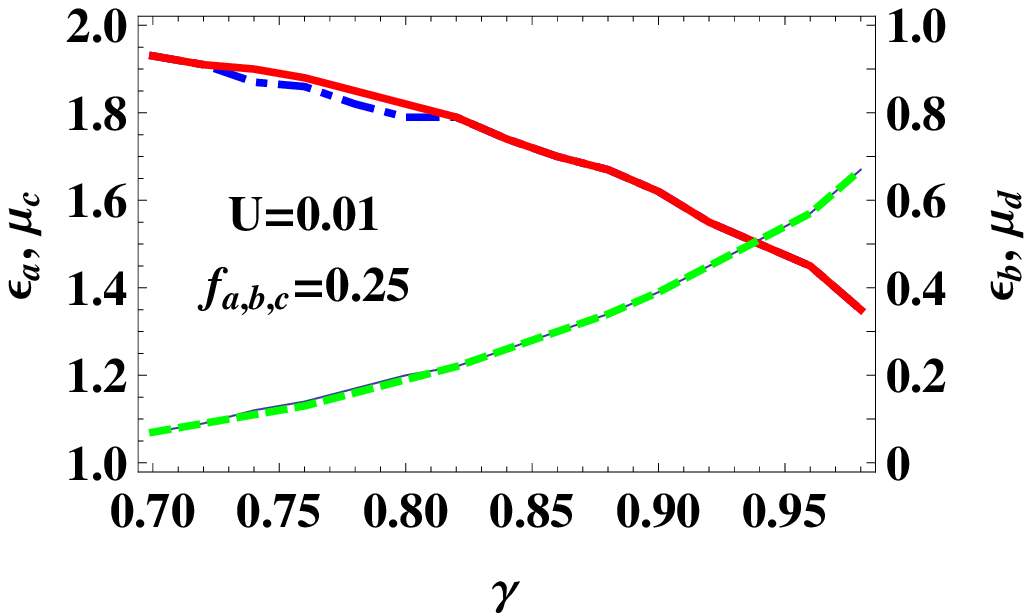}\\
\includegraphics[width=4.9in]{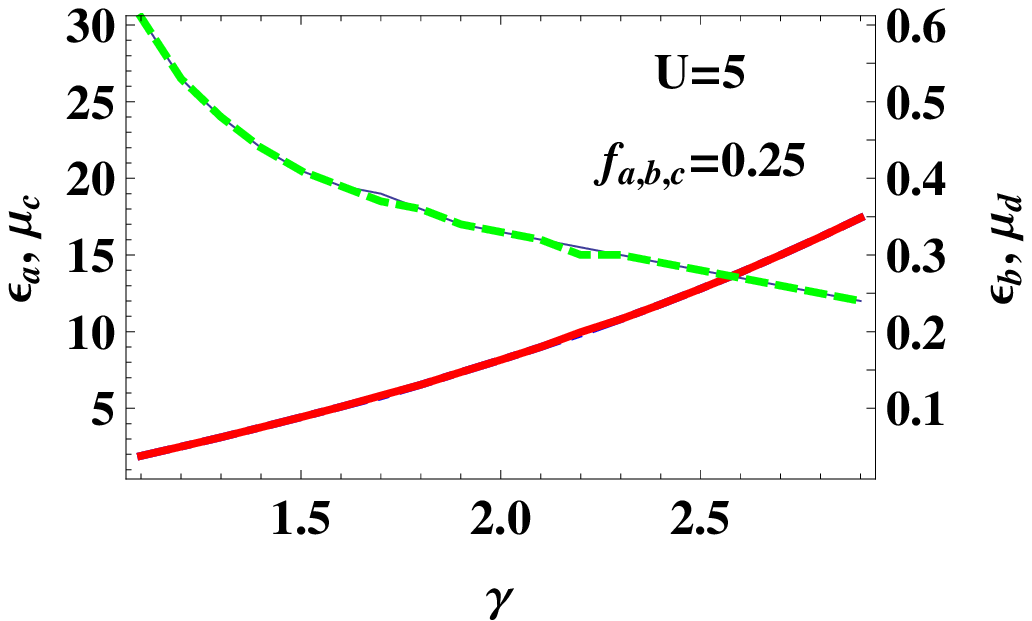}
 \caption{\label{fig5}
The  relative permittivities $\epsilon_{a}  $ (thick solid, red),
$\epsilon_{b}$ (dashed, green)
 and relative permeabilities $\mu_{c} $ (broken dashed, blue), $\mu_{d}$ (thin solid, blue)
 plotted versus $\gamma$.
The shape parameter $U=U_{a,b,c,d} = 0.01$ and volume fractions
$f_{a,b,c} = 0.25$ for $0.7 < \gamma < 1$
 (upper);
  and $U=U_{a,b,c,d} = 5$ and
$f_{a,b,c} = 0.25$ for $1.1 < \gamma < 2.9$
  (lower). }
\end{figure}

Lastly, let us comment on the restriction of the inverse
homogenization approach to those regions of spacetime which lie
outside the event horizon for Schwarzschild  spacetime and inside
the event horizon for de Sitter spacetime (there being no such
restrictions for anti-de Sitter spacetime). Equivalently, only Tamm
mediums with positive-definite relative permittivity and relative
permeability dyadics  can be realized using this approach.  The
constitutive dyadics for the Tamm medium representing
Schwarzschild-(anti-)de Sitter spacetime cannot possibly be negative
definite but, as illustrated in Figs.~\ref{F_gamma1} and
\ref{F_gamma2}, the indefinite scenario can arise  (except in the
case of  anti-de Sitter spacetime). In principle,
  metamaterials with indefinite constitutive
dyadics could be harnessed to construct a Tamm medium for the
$\gamma < 0$ regime. Indeed, experimental studies of anisotropic
dielectric \c{Eps_indefinite}  and anisotropic magnetic
\c{Mu_Indefinite1,Mu_Indefinite2} metamaterials with indefinite
constitutive dyadics have already  been reported.
 However, there are
 likely to
be major practical difficulties in using such metamaterials to
construct a $\gamma < 0 $ Tamm medium: First, not only are the
relative permittivity dyadic and the relative  permeability dyadic
both required to be
 indefinite (since the Tamm medium is an anisotropic
dielectric-magnetic medium), but these two  dyadics are required to
be identical.
 Second, the metamaterials
should be approximately nondissipative in order to faithfully
represent the Tamm medium, but  dissipation has proved to be a
severe hindrance for many metamaterials. Finally, we add that
conventional depolarization-dyadic-based homogenization techniques
(as typified by the Bruggeman formalism) cannot be used to realize
such metamaterials as nondissipative HCMs, since the indefinite
nature of the constitutive dyadics renders the depolarization
dyadics undefined \c{Michel}. This difficulty could be sidestepped
by the incorporation of a small amount of dissipation \c{MLD_MOTL},
but at the cost of compromising the extent to which the resulting
HCM represents the desired Tamm medium.

\vspace{20mm}

\noindent {\bf Acknowledgment:}  AL thanks the Charles Godfrey
Binder Endowment at Penn State for partial financial support of his
research activities.


\newpage

\newpage

\newpage

\newpage

\end{document}